\documentclass[12pt]{spieman}  
\usepackage{amsmath,amsfonts,amssymb}
\usepackage{graphicx}
\usepackage{setspace}
\usepackage{tocloft}

\title{Starshade Exoplanet Data Challenge: What We Learned}

\usepackage{fancyhdr}
\author[a]{Mario Damiano}
\author[a]{Stuart Shaklan}
\author[a,b,*]{Renyu Hu}
\author[c]{Brian Dunne}
\author[d]{Angelle Tanner}
\author[e]{Aly Nida}
\author[e]{Joseph C. Carson}
\author[a,f]{Sergi R. Hildebrandt}
\author[a]{Doug Lisman}
\affil[a]{Jet Propulsion Laboratory, California Institute of Technology, Pasadena, CA 91109, USA}
\affil[b]{Division of Geological and Planetary Sciences, California Institute of Technology, Pasadena, CA 91125, USA}
\affil[c]{Quartus Engineering, CA 92121, USA}
\affil[d]{Mississippi State University, Department of Physics and Astronomy, MS 39762, USA}
\affil[e]{College of Charleston, Department of Physics an Astronomy, SC 29424 USA}
\affil[f]{Division of Physics, Mathematics and Astronomy, California Institute of Technology, Pasadena, CA 91125, USA}

\cftpagenumbersoff{figure}
\cftpagenumbersoff{table}
\begin{document} 
\maketitle

\begin{abstract}
Starshade is one of the technologies that will enable the observation and characterization of small planets around nearby stars through direct imaging. Extensive models have been developed to describe a Starshade's optical performance and the resulting noise budget in exoplanet imaging. The Starshade Exoplanetary Data Challenge (SEDC) was designed to validate this noise budget and evaluate the capabilities of image-processing techniques, by inviting community participating teams to analyze $>1000$ simulated images of hypothetical exoplanetary systems observed through a starshade. Because the starshade would suppress the starlight so well, the dominant noise source and the main challenge for the planet detection becomes the exozodiacal disks and their structures.
In this paper, we summarize the techniques used by the participating teams and compare their findings with the truth. With an independent component analysis to remove the background, about 70\% of the inner planets (close to the inner working angle) have been detected and $\sim$40\% of the outer planet (fainter than the inner counterparts) have been identified. Also, the inclination of the exozodiacal disk can be inferred from individual images. Planet detection becomes more difficult in the cases of higher disk inclination, as the false negative and false positive counts increase. Interestingly, we found little difference in the planet detection ability between $10^{-10}$ and $10^{-9}$ instrument contrast, confirming that the dominant limitations are from the astrophysical background and not due to the performance of the starshade. Finally, we find that a non-parametric background calibration scheme, such as the independent component analysis reported here, results in a mean residual of $10\%$ the background brightness. This background estimation error leads to substantial false positives and negatives and systematic bias in the planet flux estimation, and should be included in the estimation of the planet detection signal-to-noise ratio for imaging using a starshade and also a coronagraph that delivers exozodi-limited imaging.
The results of the SEDC corroborate the starshade noise budget with realistic images, and provide new insight into background calibration that will be useful for anticipating the science capabilities of future high-contrast imaging space missions.
\end{abstract}


{\noindent \footnotesize\textbf{*}Corresponding author mail :  \linkable{renyu.hu@jpl.nasa.gov} }

\begin{spacing}{2}   

\section{Introduction}
\label{sec:intro}  

The characterization of atmospheres, climate, and potential habitability of extrasolar planets (exoplanets) with spectroscopy is at the forefront of our endeavour to explore the Universe. While short-period planets may be within the reach of the James Webb Space Telescope (JWST) for atmospheric studies through the transit technique \cite{tsai2023photochemically,grant2023jwst,madhusudhan2023carbon,hu2024secondary}, it would be challenging to characterize long-period and small planets with a secondary atmosphere (i.e., temperate and rocky planets that have N$_2$ or CO$_2$ atmospheres, or potentially habitable planets that may resemble Earth) using the transit method \cite{lim2023atmospheric}. Instead, high-contrast imaging is a recognized avenue to enable the spectroscopic studies of long-period (and thus temperate) and rocky exoplanets of nearby stars. The level of instrument contrast required to image an Earth-like planet directly around a Sun-like star ($\sim10^{-10}$) has been achieved in laboratories using a starshade \cite{Harness2021} and should be within the reach of a coronagraph \cite{Trauger2007,mennesson2024current}.

To realize the long-term vision to image and characterize Earth-like planets of nearby stars, NASA chartered the Starshade Technology Development Activity to TRL5 (S5\footnote{https://exoplanets.nasa.gov/exep/technology/starshade/}) to systematically mature the technology and close technology gaps in optical performance, formation flying, and mechanical precision and stability. S5 spearheaded a Science and Industry Partnership (SIP) to engage the broader science and technology communities during the activity. As a key outcome of this partnership, the Starshade Exoplanet Data Challenge (SEDC) \cite{Hu2021} was designed to study and validate the flow down of requirements from science capabilities to key performance parameters based on synthetic images, and to prepare the statistical tools useful for the analysis of data that involve the starshade technology. An Earth-sized planet in the habitable zone of a nearby star would likely appear to be dimmer than the starlight reflected by the exozodiacal dust \cite{Hu2021b,kammerer2022simulating,currie2023mitigating} and, in some cases, the sunlight scattered by the edge of the starshade (referred to as the ``solar glint'') \cite{Hu2021b}. These noise sources and the planet overlap in the image and are not easily separable, and thus it is essential to use synthetic images to test planet detection capabilities.

The objectives of the SEDC is (1) to validate requirements from science to key performance parameters, (2) to quantify the accuracy of calibration of solar glint and exozodiacal light, and (3) to prepare the science community for analyzing starshade exoplanet observations. To support the SEDC, the S5 team has simulated 2880 images to explore relevant star and planet parameters, exozodiacal disk density and orientation, as well as the wavelength, instrumental noise level, and integration time (see Ref.~\citenum{Hu2021} for details). There are 10 star-planet scenarios in total (see Tab. \ref{tab:scenarios}), and for each of them, three levels of exozodiacal disk density are explored (1, 3, and 10 zodis, except for $\tau$~Ceti that was 3, 10, and 30 zodis), with and without density clumps in resonance with the embedded planets \cite{stark2011transit}. We assume that the observation is performed with a 2.4-m space telescope (which determines the point spread function) coupled with a starshade at two broad wavelength bands. The inner working angle (IWA) of the starshade is 72 mas at 415-552 nm and 104 mas at 615-850 nm. We consider a standard starshade that delivers $10^{-10}$ instrument contrast at the IWA and also a degraded starshade that delivers $10^{-9}$, resulting in two sets of images. We provide two images for each setup, where the locations of the planets are different between the two images. The two images are however not time-stamped and thus cannot be associated via Keplerian motions. We pick the integration time so that the inner planet would have an idealized S/N of 5, 10, or 20. The ``idealized S/N'' is the planet's S/N for the perfect background subtraction to the photon-noise limit. All images are simulated using the Starshade Imaging Simulation Toolkit for Exoplanet Reconnaissance (SISTER) \cite{hildebrandt2021starshade}, and it is the first time that the optical effects of the telescope jitter and the formation flying of the starshade have been included in synthesized images. The images were released on April 7$^{th}$, 2021 and remain available for download on a dedicated webpage for the SEDC \footnote{https://exoplanets.nasa.gov/exep/technology/starshade-data-challenge/}. Two funded participating teams were selected based on proposals, while the data challenge is open to the broad community. The true values have been posted onto the same webpage after receiving the reports from the two participating teams, and we encourage continuing exploration of the post-processing techniques using this unique dataset.

\begin{table}
	\center
	\caption{Simulated planet-star system scenarios in SEDC \cite{Hu2021}. Except for $\tau$~Ceti, the inner planet is located at an orbital distance that is equivalent (in terms of the total radiation received by the planet) to 1 AU in the solar system, and the outer planet is located at at an orbital distance that is equivalent to 1.5 AU in the solar system. For $\tau$~Ceti we use the two long-period planets (planets e and f) suggested by radial-velocity surveys \cite{feng2017color} as the inner and outer planets, and add a hypothetical, 1-$R_{\oplus}$ planet at the orbital distance that is the geometric mean of the values for inner and outer planets. Next to the planet radius, in parentheses, we report the average background/planet flux ratio for the lowest exozodi level considered for the scenario, i.e., 1 zodi for scenarios 3 -- 10 and 3 zodis for scenarios 1 and 2.}
	\label{tab:scenarios}
	\begin{tabular}{llllll}
		
		\hline \hline
        ID & Star & Disk Inc. [deg] & Planet 1 Rad. [$R_{\oplus}$] & Planet 2 Rad. [$R_{\oplus}$] & Planet 3 Rad. [$R_{\oplus}$] \\
        \hline
        1 & $\tau$ Ceti & 35 & 1.65 (3.70) & 1.0 (57.50) & 1.65 (7.55)\\
        2 & $\tau$ Ceti & 35 & 2.05 (1.86) & 1.0 (58.01) & 2.05 (3.55) \\
		3 & $\epsilon$ Indi A & 30 & 1.0 (20.26) & 1.0 (35.25) & N/A \\
		4 & $\epsilon$ Indi A & 80 & 1.0 (26.29) & 1.0 (49.38) & N/A \\
		5 & $\sigma$ Draconis & 30 & 1.6 (2.56) & 1.6 (4.20) & N/A \\
		6 & $\sigma$ Draconis & 80 & 1.6 (3.40) & 1.6 (6.26) & N/A \\
		7 & $\sigma$ Draconis & 30 & 2.4 (0.87) & 2.4 (1.42) & N/A \\
		8 & $\sigma$ Draconis & 80 & 2.4 (1.17) & 2.4 (2.14) & N/A \\
		9 & $\beta$ CVn & 30 & 2.4 (1.63) & 2.4 (2.46) & N/A \\
		10 & $\beta$ CVn & 80 & 2.4 (2.22) & 2.4 (4.30) & N/A \\
		\hline
	\end{tabular}
\end{table}

Whether background subtraction can be achieved close to the photon-noise limit has a strong bearing on forecasting the capability of future facilities to detect exoplanets. This is because the exozodiacal disk, and to a lesser extent solar glint from the starshade, is expected to be bright with respect to a small ($1-4$ $R_{\oplus}$) planet in the habitable zone of nearby stars \cite{Hu2021b}. 
Existing mission yield studies\cite{stark2014maximizing,stark2019exoearth,Morgan2019,Morgan2021} and the HabEx and LUVOIR final reports \cite{Gaudi2020,2019arXiv191206219T} generally assumed background calibration to the photon-noise limit. In the context of exoplanet direct imaging using a coronagraph, the stability of the residual starlight (i.e., the ``speckles'') leads to imperfect background calibration and a potentially irreducible noise floor \cite{nemati2020method,mennesson2024current}. When using a starshade, the leaked starlight and its variability becomes generally negligible, especially when the angular separation is greater than the IWA \cite{Hu2021b}. Recently, Refs.~\citenum{kammerer2022simulating,currie2023mitigating} studied the data analysis techniques of the simulated images using a coronograph and suggested that the background from the exozodiacal disk can be subtracted to close to the photon-noise limit, depending on the disk density, inclination, and whether the disk has resonant structures.
Refs.~\citenum{kammerer2022simulating,currie2023mitigating} however did not perform ``blind'' experiments, i.e., the same team simulated the images and analyzed them.
Here, with SEDC, we aim to provide an empirical constraint on the precision of background calibration based on independent data analyses performed by the participating teams, who had no knowledge of the astrophysical scenarios when analyzing the images. 

While this study assumes a starshade as the starlight suppression instrument, we expect the findings to be generally applicable to the situations where the residual starlight is negligible in the overall noise budget. While the synthesized images in this study fully include the starshade-specific noise terms -- solar glint, reflected Earthshine, micrometeorite damages, and their variability due to the formation flying motion of the shade -- it has become clear a posteriori that none of these terms is likely dominant in the background when searching for Earth-like planets around nearby stars. This study can thus be regarded, more broadly, as path-finding for the search for planets in the exozodi-limited regime regardless of the starlight suppression system. The applicability thus includes the search for Earth-like exoplanets using a coronagraph on the Habitable Worlds Observatory, assuming that the coronagraph would have an instrument contrast at the $\sim10^{-10}$ level and a speckle field stability that does not become the bottleneck for planet detection \cite{stark2015lower,mennesson2024current}.

In this paper we describe the image analysis methods developed by the two participating teams (Sec.~\ref{sec:summary}) and compare their findings with the truths (Sec.~\ref{sec:eval}). The planet coordinates and brightness are unknown to the participating teams. By comparing the reported planet detection/fluxes to the truths with a heuristic model of background estimation and planet detection, we provide an evaluation for the precision of background estimation achieved by the participating teams (Sec.~\ref{sec:heuristic}). Based on this, we discuss the high-level implication in Sec.~\ref{sec:discussion} and conclude with anticipation of future steps in Sec.~\ref{sec:conclusion}.

\section{Summary of Participating Team Analyses}
\label{sec:summary}

Here we provide a synopsis of the approaches taken by the two participating teams. The full reports provided by the participants are available for download from the SEDC webpage.

\subsection{Quartus Engineering}
\label{sec:quartus}

In general, a given simulated image includes relatively faint signal from exoplanets, in addition to background signal from various noise sources: residual starlight, solar glint and other stray light sources, exozodiacal light, detector noise, and variability resulting from the starshade’s motion and telescope jitter. The Quartus Engineering team broke the image analysis problem into the following sub-problems: background estimation, planet detection, and planet parameter estimation. The background estimation problem proved to be the most challenging sub-problem, and the Quartus Engineering team explored both parametric and non-parametric approaches to address this sub-problem. The parametric approach attempts to explicitly model the backgrounds, find best fit parameters, and then subtract the best fit background model. The non-parametric category includes approaches which do not use an explicit astrophysical or otherwise explicit mathematical model of observed backgrounds, and instead attempt to exploit underlying differences in the behavior of planet signals versus background signals when considering a large training set. Specifically, two non-parametric approaches were explored: Principal Component Analysis and Independent Component Analysis (PCA and ICA). These approaches attempt to find a low-dimensional latent representation of the background signals.

\subsubsection{Parametric approach for background estimation}

The parametric background model developed consisted of an assumption of an exozodiacal disk which is circular, with a 1D function describing a radially symmetric intensity profile. Parameters for inclination and orientation relative to the imaging system are also included. This can be considered an ``empirical model'' in the sense that it is not derived from astrophysical principles, and is instead intuited from inspection of the image data without any astrophysical reasoning. The model has the following parameters:
\begin{itemize}
    \item Center position $x_c$, $y_c$ (pixels)
    \item Inclination and orientation $i$, $\theta$ (deg)
    \item 1D function for symmetric intensity roll-off:
    \begin{itemize}
        \item f(r)= $se$$^{(ar^{2}+br+c)}$, where:
        \begin{itemize}
            \item $r$ is radial distance from the center
            \item $s$ is an intensity scale factor
            \item $a$, $b$, and $c$ are polynomial coefficients
        \end{itemize}
    \end{itemize}
\end{itemize}
This gives a total of 8 parameters, denoted by parameter vector $p$: $p = [x_c, y_c, i, \theta, s, a, b, c]$. Although this model is not derived from astrophysical principles, it has some physical interpretability in the sense that $x_c$, $y_c$ represent estimated disk center, while $i$, $\theta$ represent disk inclination and orientation, under the rough assumption of an infinitely thin disk with perfect radial symmetry.

We find that this approach generally gave reasonable success for removing smooth disk scenarios. However, it performed worse in those scenarios where the thickness of the disk is not negligible or the structure of the disk could not be described as smooth.

\subsubsection{Non-parametric approach for background estimation}

Developing a parametric model with both sufficient fidelity to capture all the background effects and also the right properties for an inverse estimation problem appeared unrealistic given the diversity of the disk density and morphology among the simulated images. Therefore, the Quartus Engineering team switched focus to non-parametric background estimation approaches, since there are a few reasons to believe that the planetary signal can be separated from background signal due to underlying differences in their behavior across a large dataset:
\begin{itemize}
    \item Planets move between multi-epoch observations, while smooth, axisymmetric exozodiacal backgrounds do not, relative to the star;
    \item The shape and scale of the disk structure in the image changes significantly as a function of wavelength due to the different size of the PSF, whereas planet positions do not. This is relevant when combining images from the same observation into a multi-channel image considered as a single sample;
    \item Disk structures have underlying similarity between different scenarios and dominate the signal energy;
    \item Planet signals are point sources with low signal energy, appearing in non-repeatable positions.
\end{itemize}
A general idea to exploit these underlying effects to separate planets from background is to use dimensionality reduction techniques to find a latent model which captures backgrounds but not planets. In essence, a low dimensionality latent model will have a de-noising effect, where the planet signal can be thought of as a noise signal on top of the broad background structure.

\begin{figure}[!h]
\centering
		\includegraphics[width=\columnwidth]{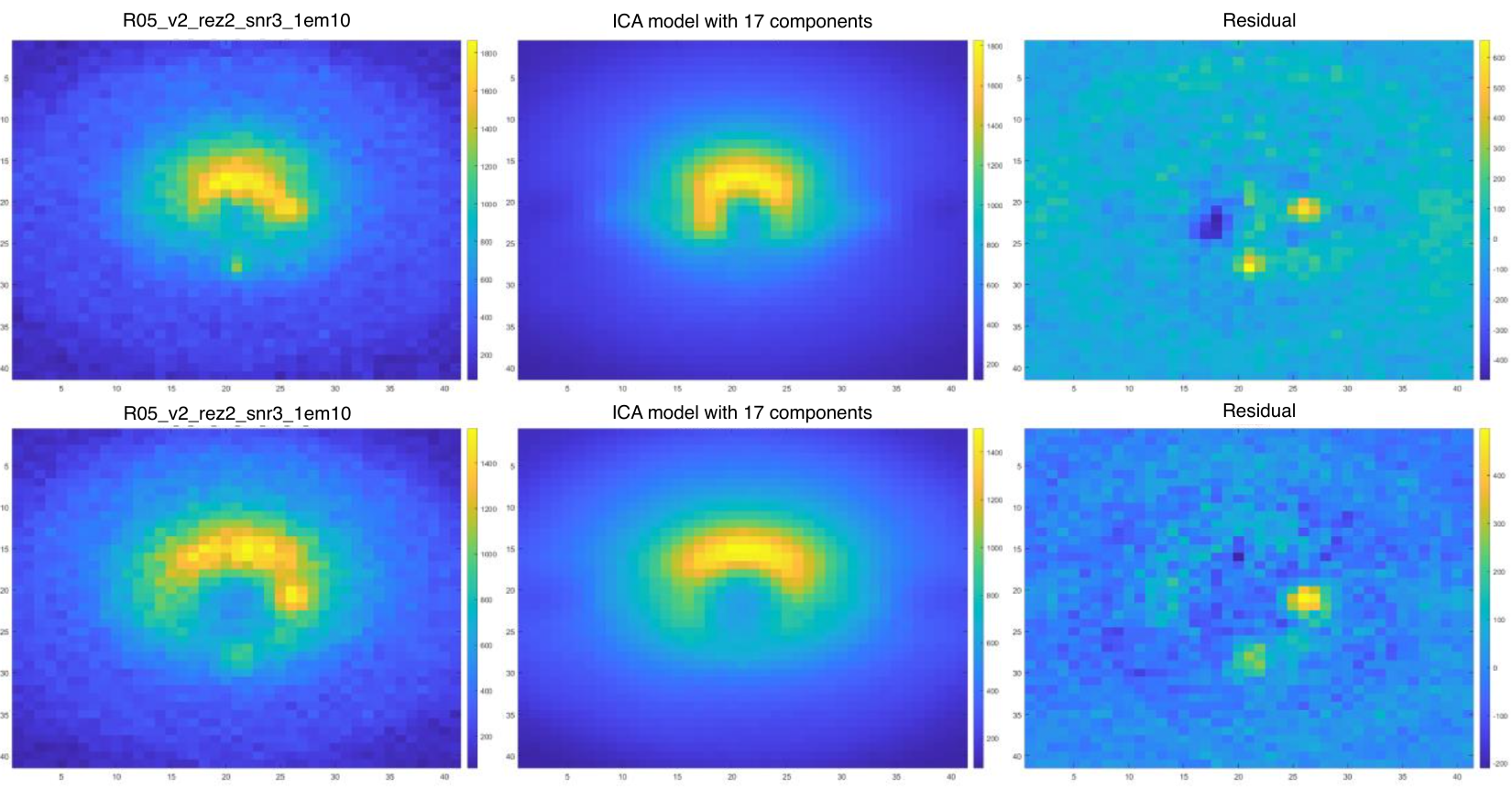}
		\caption{Example of ICA-based background subtraction results for ``\texttt{R05\_v2\_rez2\_snr3\_1em10}'' scenario from the Quartus Engineering analysis. A $2\times3$ set of subplots is shown where rows 1 and 2 correspond to passbands [425-522] and [615-800] respectively. Columns 1, 2, and 3 of the plot correspond to the original image, the estimated background, and the residual, respectively. \label{fig:ICA_decomp}}
\end{figure}

The Quartus Engineering team thus explored PCA and ICA as potential tool for non-parametric decomposition. In both cases, the image data is collected into a matrix, where the rows represent samples, and the columns represent variables, which in this case are the image pixels flattened into a row. ICA could be seen as a generalization of PCA, as the linear transformation from the original space to the rotated space in which the data variance is maximised and is not required to be orthonormal. Therefore, the Quartus Engineering team eventually adopted ICA to perform the background subtraction and planetary signal extraction. An application of ICA to a SEDC image is shown in Fig.~\ref{fig:ICA_decomp}.

\subsubsection{Planet detection and estimation}

Following the background subtraction, the planet detection step starts with the background subtracted image, which is denoted as I$_f$ (foreground image). In the current workflow I$_f$ = I$_o$ \text{-} I$_r$, where I$_o$ is the observed image, and I$_r$ is the image obtained by using the ICA model to project onto the latent space, followed by reconstruction back into image space. For example, the residual images of the ICA processing (e.g., column 3 of Fig.~\ref{fig:ICA_decomp}) are considered to be foreground images for the purpose of planet detection.

The pipeline implemented for the purposes of the data challenge includes the following steps:
\begin{itemize}
    \item Matched filtering: The foreground images are filtered with the appropriate PSF as a function of wavelength and instrument scenario;
    \item Local maxima detection: local maxima detection is performed on the matched filter output to find regions of the image which are well matched to the PSF, indicating possible presence of a point source;
    \item Outlier detection: given the distribution of local maxima intensities found in the previous step, uni-variate outlier detection is performed to detect which local maxima do not match the main distribution. The Grubbs test is used to detect local maxima which are outliers in terms of matched filter output intensity;
    \item Keep the N most intense positive outlier detections and consider them as planet detections.
\end{itemize}

\begin{figure}[!h]
\centering
		\includegraphics[width=\columnwidth]{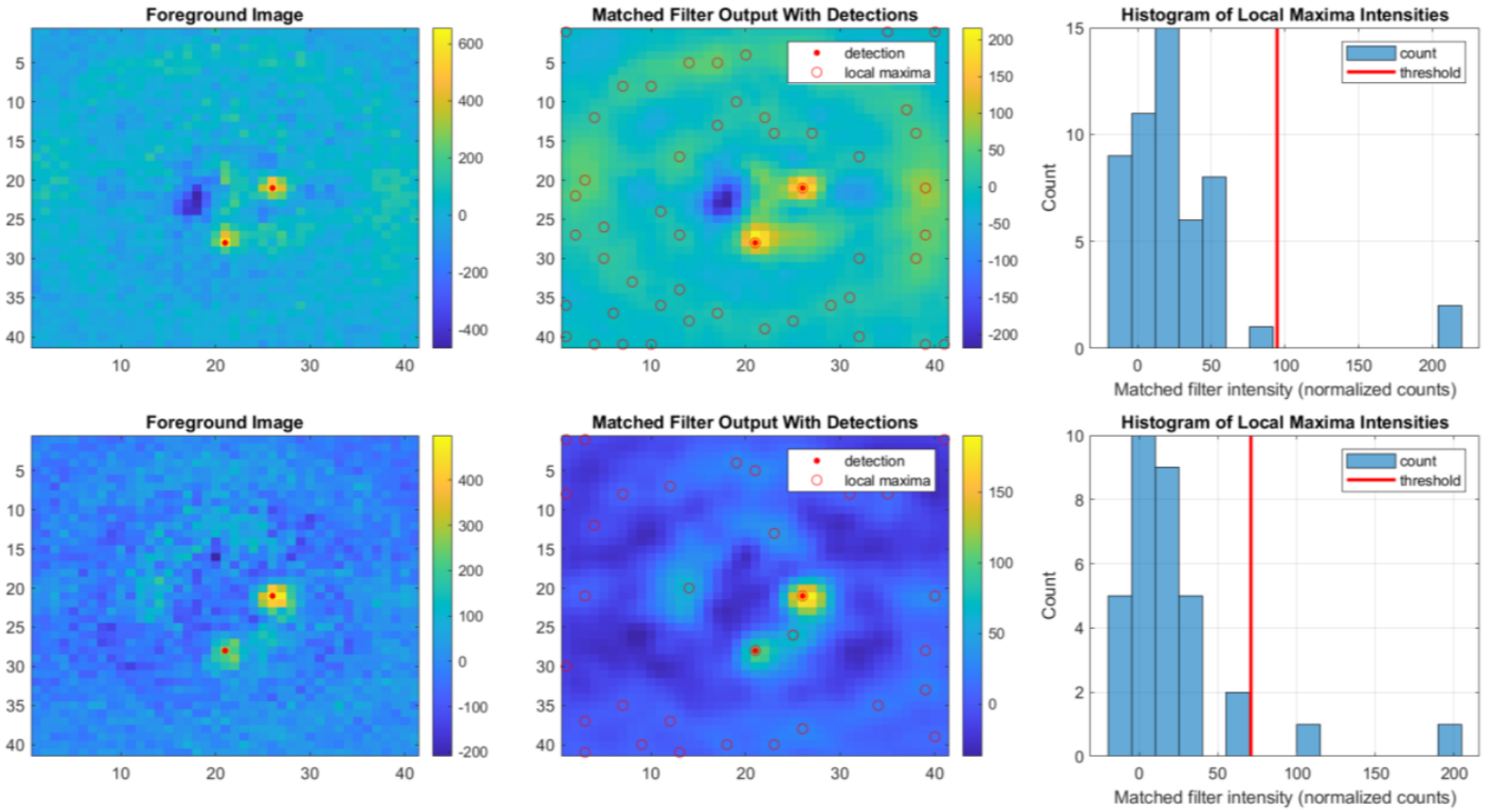}
		\caption{Example planet detection results for the ``\texttt{R05\_v2\_rez2\_snr3\_1em10}'' scenario from the Quartus Engineering analysis. A $2\times3$ set of subplots is shown where rows 1, 2 correspond to passbands [425-522] and [615-800] respectively. Columns 1, 2, and 3 of the plot correspond to: foreground images with detections, matched filter images with detections and local maxima, and a histogram of local maxima intensities with outlier/detection threshold. \label{fig:planet_detection}}
\end{figure}

A visualization of detection results for the ``\texttt{R05\_v2\_rez2\_snr3\_1em10}'' scenario is shown in Fig.~\ref{fig:planet_detection}. It is apparent in Fig.~\ref{fig:planet_detection} that, while the automated detection is successful for two planets, imperfect background subtraction is affecting the background level, creating valleys with no local maxima, potentially leading to false negatives. Also, the threshold used in the planet detection scheme may lead to the identification of false positives that can be mistakenly identified as planets. The choice of the threshold acts as a trade off between sensitivity and specificity. The method is, therefore, identifying point-like signals, which includes both planets and certain disk structures. For many images, imperfect background subtraction appears to be the limiting factor.

While the detection problem described previously can be summarized as detecting pixels in the image which correspond to planets, the problem of estimating parameters for each detected planet can be considered as an independent sub-problem. The following approach was taken for estimating parameters per planet:
\begin{itemize}
    \item For each planet detection in a given image, select an $N\times N$ pixel region of interest (ROI) centered on the detection. In this case the selected ROI width was 7 pixels, covering the PSF in all scenarios;
    \item Fit a parameterized model of a PSF to the sampled ROI region, using a standard nonlinear least squares approach;
    \item Use the center location and magnitude from the best-fit PSF to ultimately compute the planet location and uncertainty, photon counts, and planetary signal to noise ratio (S/N).
\end{itemize}

\begin{figure}[!h]
\centering
		\includegraphics[width=\columnwidth]{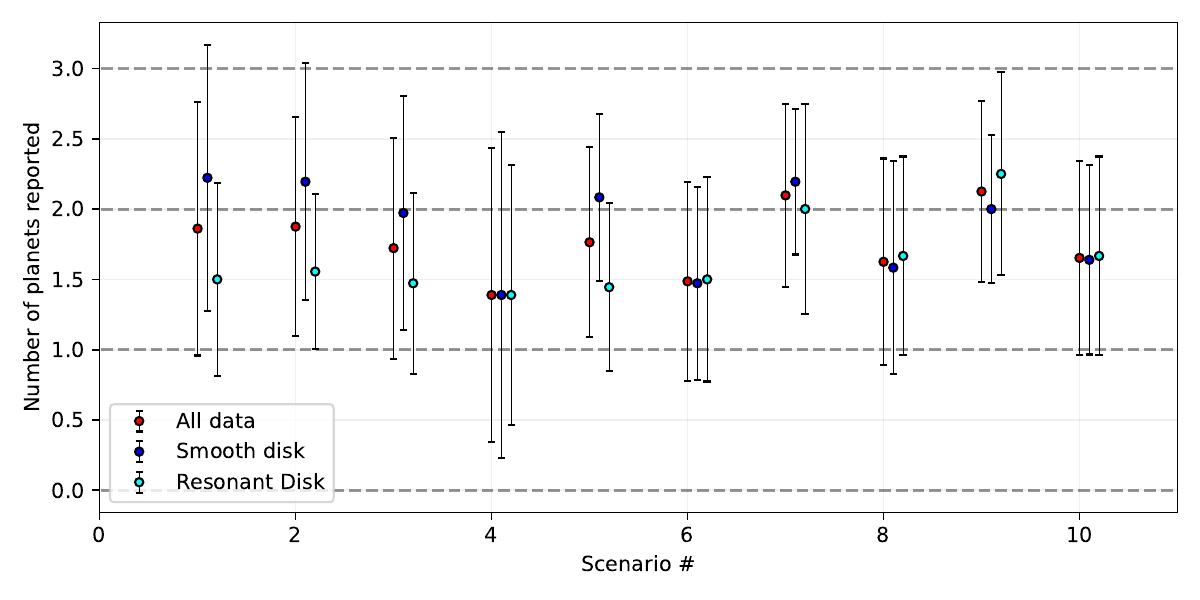}
		\caption{Mean and standard deviation of the number of planets reported by the Quartus Engineering team per image for each star-planet scenario. The scenario \# corresponds to Table~\ref{tab:scenarios}. The smooth and resonant cases are also tallied separately. \label{fig:planet_reported}}
\end{figure}

By applying the algorithm to all synthesized images, the Quartus Engineering team was able to isolate the planetary signal, extract it, and estimate the planetary S/N. Across the 10 star-planet scenarios, in average between one and two planets were reported per image as shown in Fig.~\ref{fig:planet_reported}. Note that the number of reported planets does not significantly vary between the smooth and resonant exozodiacal disk cases except for scenarios number 1, 2, 3, and 5 (i.e., $\tau$~Ceti and face-on cases of $\epsilon$~Indi~A and $\sigma$~Draconis), in which more planets are reported for the smooth disk cases.

\subsection{Mississippi State University}
\label{sec:mississippi}

The Mississippi State University team utilized three main techniques to characterize and then remove light from the image which is not associated with the planet. The techniques employed include: (1) PSF subtraction, (2) disk modeling, and (3) multi-epoch differential imaging.

PSF subtraction involves subtracting a model or PSF in the absence of any apparent exozodical emission to better reveal the faint emission from the planetary point source. Disk modeling is used to create a simple model of any circumstellar emission present in the image for a single epoch. Once the model is made, it is subtracted from the science image leaving emission from the planet. Multi-epoch subtraction is the method of using two different epochs where it is assumed that the planet has changed its position angle while any circumstellar disk emission remains stable. To characterize what type of analysis is needed to be performed on each image, an initial visual inspection was performed. For those images which appeared to not have any extended emission near the inner working angle, the Mississippi State University team performed an additional analysis comparing the profile of an image slice through the position of the star with a similar slice of the model centered PSF. If there was a contribution of some extended emission, this image was slated for additional disk modeling to better remove this emission prior to analyzing the emission from the planet. A more thorough determination could also be done by subtracting a PSF from the image and highlighting any significant residual emission.

$\sim$30 images have been analyzed using these techniques. The locations and planetary flux ratios of $\sim$30 planets have been reported in total.

\section{Evaluation of Participating Team Analyses}
\label{sec:eval}

\subsection{Quartus Engineering}

\subsubsection{Estimate of the inclination}

\begin{figure}[!h]
\centering
		\includegraphics[width=\columnwidth]{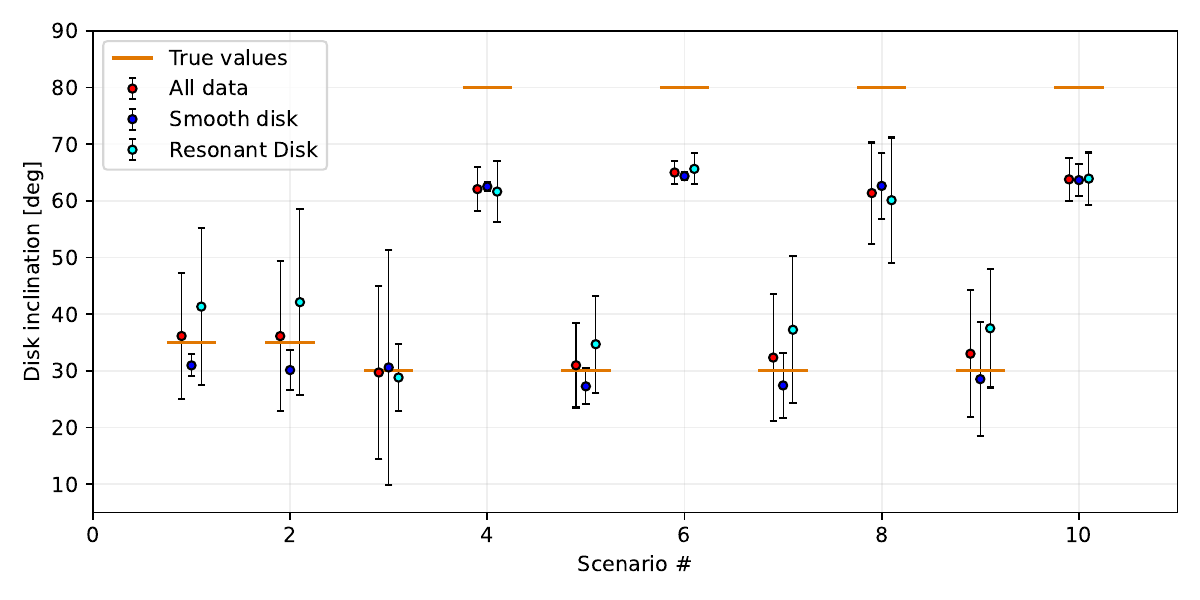}
		\caption{Disk inclination reported by the Quartus Engineering team, averaged for each of the scenarios and compared with the true values. The scenario \# corresponds to Table~\ref{tab:scenarios}. The breakdown between the smooth and resonant disk cases is also shown. \label{fig:incl_10}}
\end{figure}

The parametric background estimation method adopted by the Quartus Engineering team provides the planetary system inclination from individual images. The reported inclination values from the 10$^{-10}$ instrument contrast cases are shown in Fig.~\ref{fig:incl_10}. A clear split in two populations is shown in the data, i.e., high and low disk inclination. 
The results reported by the Quartus Engineering team agree with the truth for the low inclination cases, but it is systematically biased for the high inclination cases. This is likely due to the fact that the disk is optically thick and it is difficult to distinguish between a perfectly edge-on disk (i.e., 90$^{\circ}$ inclination) from a high-inclination disk, e.g., $\ge$60$^{\circ}$. The breakdown in smooth and resonant disk cases shows that it is more difficult (higher variance) to accurately evaluate the inclination of the system from a single image if the disk has resonant structures.

\subsubsection{True positives, false positives, and false negatives}
\label{sec:tpfpfn}

\begin{figure}[!h]
\centering
        \includegraphics[width=\columnwidth]{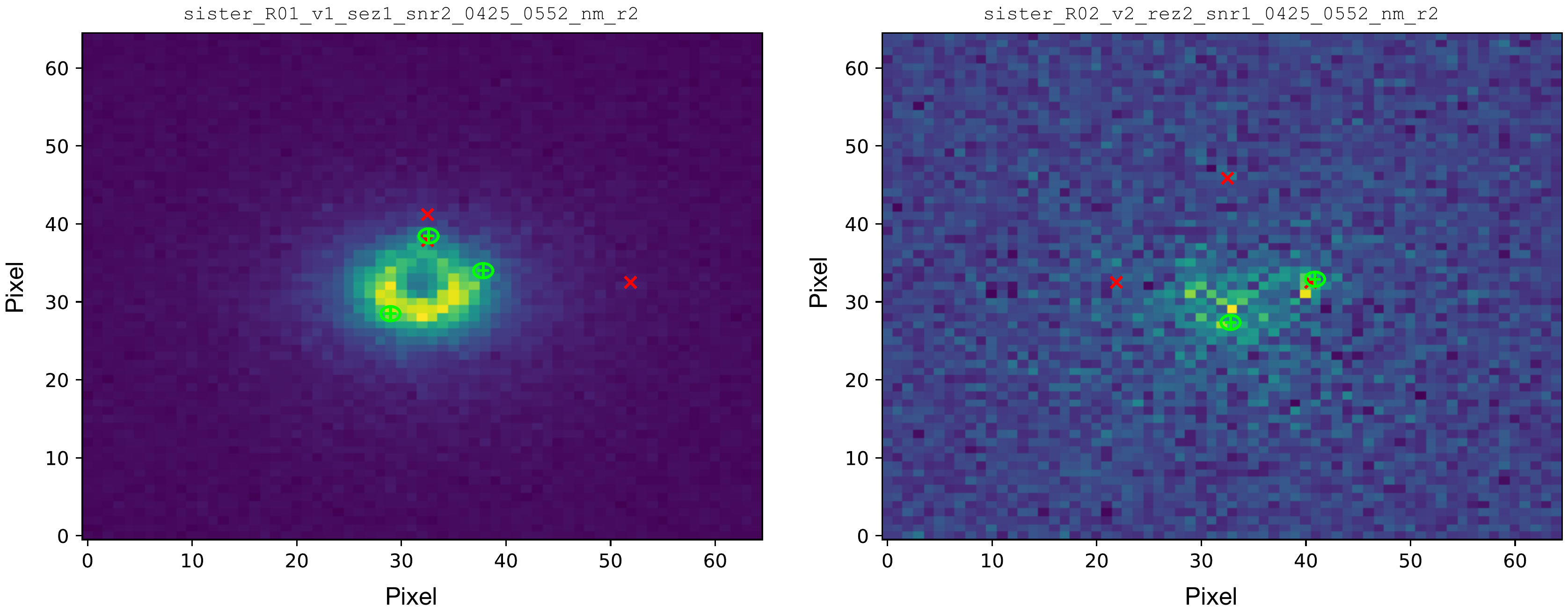}
		\caption{Two example images from the SEDC. The red crosses identify the position of the simulated planets, and the green crosses identify the position reported by Quartus Engineering. The green circles symbolize the PSF aperture used to confirm or reject a reported planet as true positive. \label{fig:img}}
\end{figure}

The Quartus Engineering team reported 1267 planets from the images that assume the instrument contrast level of 10$^{-10}$ and 1293 planets from the images that assume the instrument contrast level of 10$^{-9}$. 
To the tally of true positives, false positives, and false negatives, we define a true positive if a reported planet with its position uncertainty falls inside the simulated planet's PSF. Those reported planets that fall outside are instead considered false positives. Then, as we include two to three planets in each image (Table~\ref{tab:scenarios}), if not all planets in the image are detected, the missing planets are counted as false negatives. In Fig.~\ref{fig:img}, we show two examples to visualize the definition of true positive, false positive, and false negative. 

As shown in Table~\ref{tab:summary_stats}, approximately half of the planets in the synthesized images are detected. $70-75\%$ of the inner planets are detected but only $\sim28\%$ of the outer planets are detected. There appear to be no substantial difference on the planet detection rate between the instrument contrast level of $10^{-10}$ and $10^{-9}$, and a moderate increase in the false positive counts when the instrument contrast level is higher.

\begin{table}[]
    \centering
    \caption{Counts of detected planets, false positives, and false negatives reported by the Quartus Engineering team.}
    \begin{tabular}{lcc}
        \hline\hline
         & \multicolumn{2}{c}{\textit{Instrument contrast levels}} \\
         & 10$^{-10}$ & 10$^{-9}$ \\
         \hline
        \textit{Inner$+$outer planets} \\
        Total simulated & \multicolumn{2}{c}{1584} \\
        Planets reported & 1267 & 1293 \\
        True detected & 778 & 746 \\
        True detected rate [\%] & 49.1\% & 47.1\% \\
        False positives & 489 & 547 \\
        False negatives & 421 & 407 \\
        \hline
        \textit{Inner planets} \\
        Total simulated & \multicolumn{2}{c}{720} \\
        True detected & 535 & 507 \\
        True detected rate [\%] & 74.3\% & 70.4\% \\
        \hline
        \textit{Outer planets} \\
        Total simulated & \multicolumn{2}{c}{864} \\
        True detected & 243 & 239 \\
        True detected rate [\%] & 28.1\% & 27.7\% \\
        \hline\hline
    \end{tabular}
    \label{tab:summary_stats}
\end{table}

\begin{figure}[!h]
\centering
		\includegraphics[width=\columnwidth]{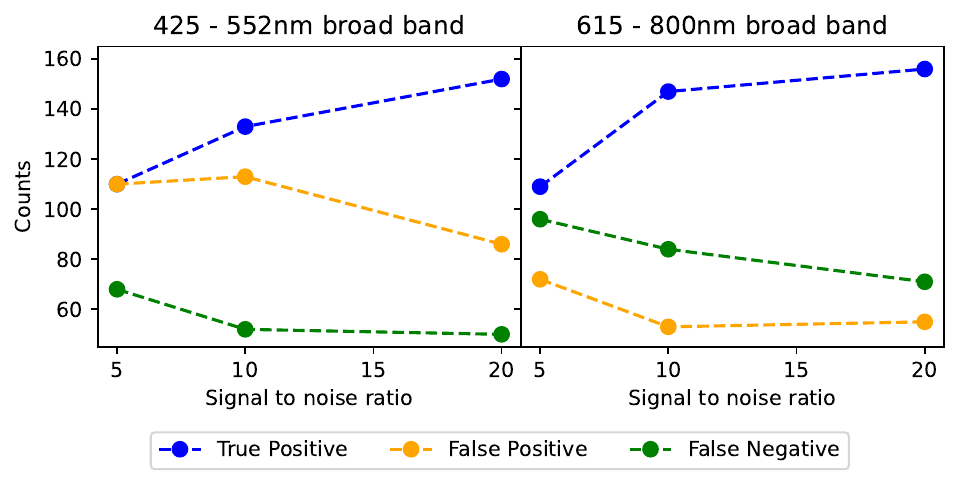}
		\caption{True positive, false positive, and false negative counts as a function of the idealized S/N for the two broadbands explored, based on the report of the Quartus Engineering team. The true positive rate is comparable between the two bands. The false positive counts are lower in the longer wavelength band and the false negative counts are lower in the shorter wavelength band. For each S/N case, there are 264 true planets in total. \label{fig:wl}}
\end{figure}

\begin{figure}[!h]
\centering
	\includegraphics[width=0.6\columnwidth]{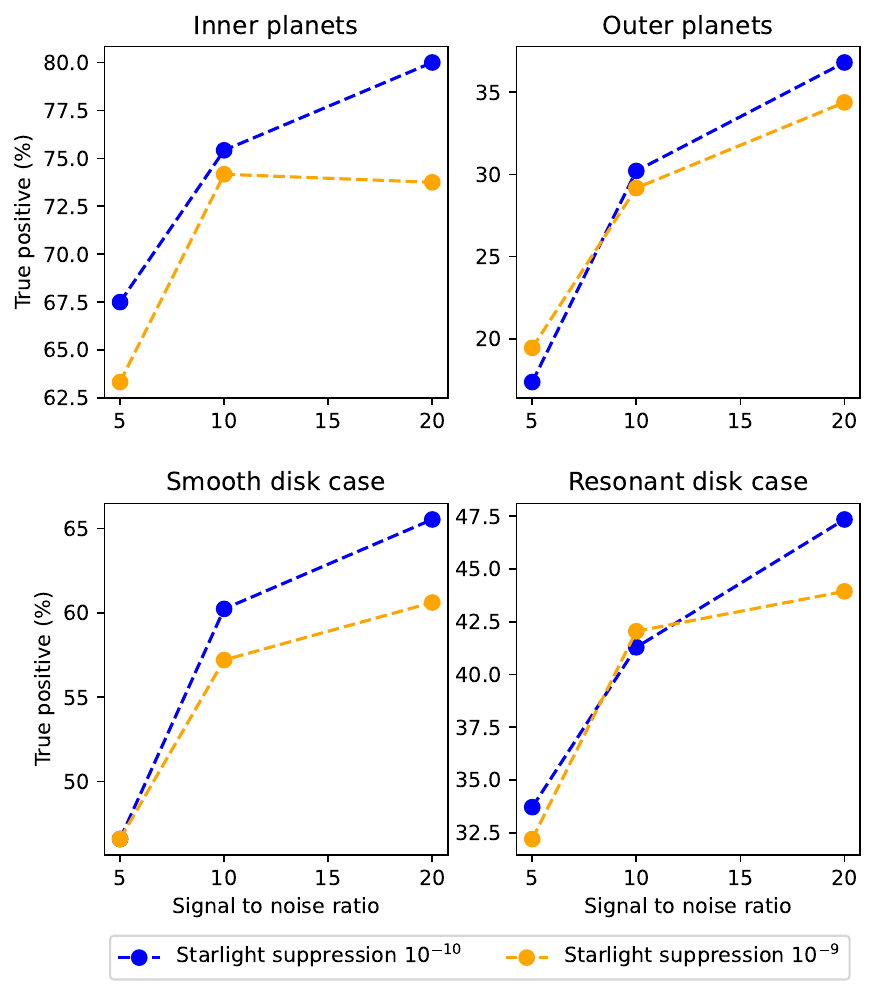}
  \caption{\textbf{Top:} True positive rates as a function of the input S/N (defined for the inner planet) for the $10^{-10}$ and $10^{-9}$ instrument contrast performance levels. \textbf{Bottom:} Breakdown of the true positive rate to detect the inner planets between the smooth disk and resonant disk cases. Based on the report of the Quartus Engineering team. }
  \label{fig:true_positive}
\end{figure}

We break down the planet detection counts by the idealized S/N, the two wavelength broadbands, and the smooth versus resonant disk cases in Figs.~\ref{fig:wl} and \ref{fig:true_positive}. The idealized S/N is defined for the inner planet. As expected, the true positive counts increase with a higher idealized S/N, and the false positive and false negative counts decrease (Fig.~\ref{fig:wl}). The gain in the planet detection rate is significant between the idealized S/N of 5 and 10, while the increase is less steep between a S/N of 10 and 20 (Fig.~\ref{fig:true_positive}). There is no obvious sensitivity on the two wavelength broadbands for planet detection.

With a higher level of leaked starlight and solar glint in the $10^{-9}$ instrument contrast cases than the $10^{-10}$ instrument contrast cases, the detection rate is only slightly lower. The difference is more evident for the inner planet and at a higher S/N (Fig.~\ref{fig:true_positive}). However, the true positive rates with a resonant disk is substantially lower than the smooth disk cases (Fig.~\ref{fig:true_positive}). These observations indicate that distinguishing the background (dominated by the exozodi disk) and the planets, rather than the instrumental background and noises, is the limiting factor for planet detection. 

\subsubsection{Origin of the false positives and false negatives}

\begin{figure}[!h]
\centering
		\includegraphics[width=\columnwidth]{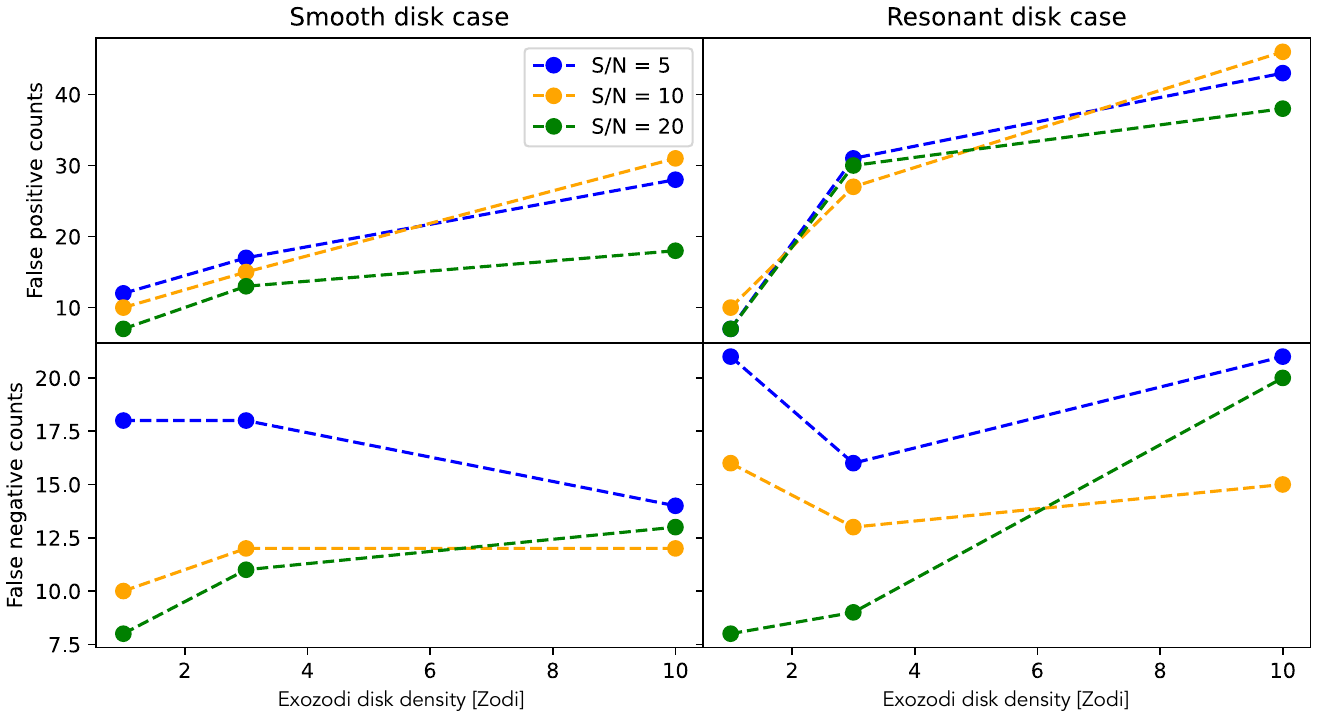}
		\caption{False positive and false negative counts as functions of the exozodiacal disk density for the smooth and resonant disk cases, reported by the Quartus Engineering team. The false positive and false negative counts generally increase with disk density for both the smooth and resonant disk cases for all the three S/N cases. Exceptionally, in the smooth disk and S/N$=$5 cases, the false negative count decreases with the exozodi level, and in the resonant-disk cases, the false negative count is the lowest for 3-zodi disk. The tally does not include Scenarios 1 and 2 as they have been simulated with different exozodi levels. \label{fig:false_positives}}
\end{figure}

We report the breakdown of the false positive cases and the false negative cases as a function of S/N and the disk density level in Fig.~\ref{fig:false_positives}. 
The false positive counts positively correlate with the exozodiacal disk density level for all of the three (idealized) S/N levels, and more false positives are found in the resonant disk cases than in the smooth disk cases. Also, increasing the S/N results in lower false positive counts in the smooth disk cases, while this trend is not obvious in the resonant disk cases. Similarly, the false negative counts are generally positively correlated with the exozodi level except for the S/N$=$5 smooth-disk cases. As S/N increases, the number of the false negatives drops as expected. These observations indicate that the exozodi disk, and particularly its clumpiness, plays a crucial role in generating the false positives and false negatives. With resonant disk structures, simply increasing the S/N does not effectively remove false positives.

\begin{figure}[!h]
\centering
		\includegraphics[width=\columnwidth]{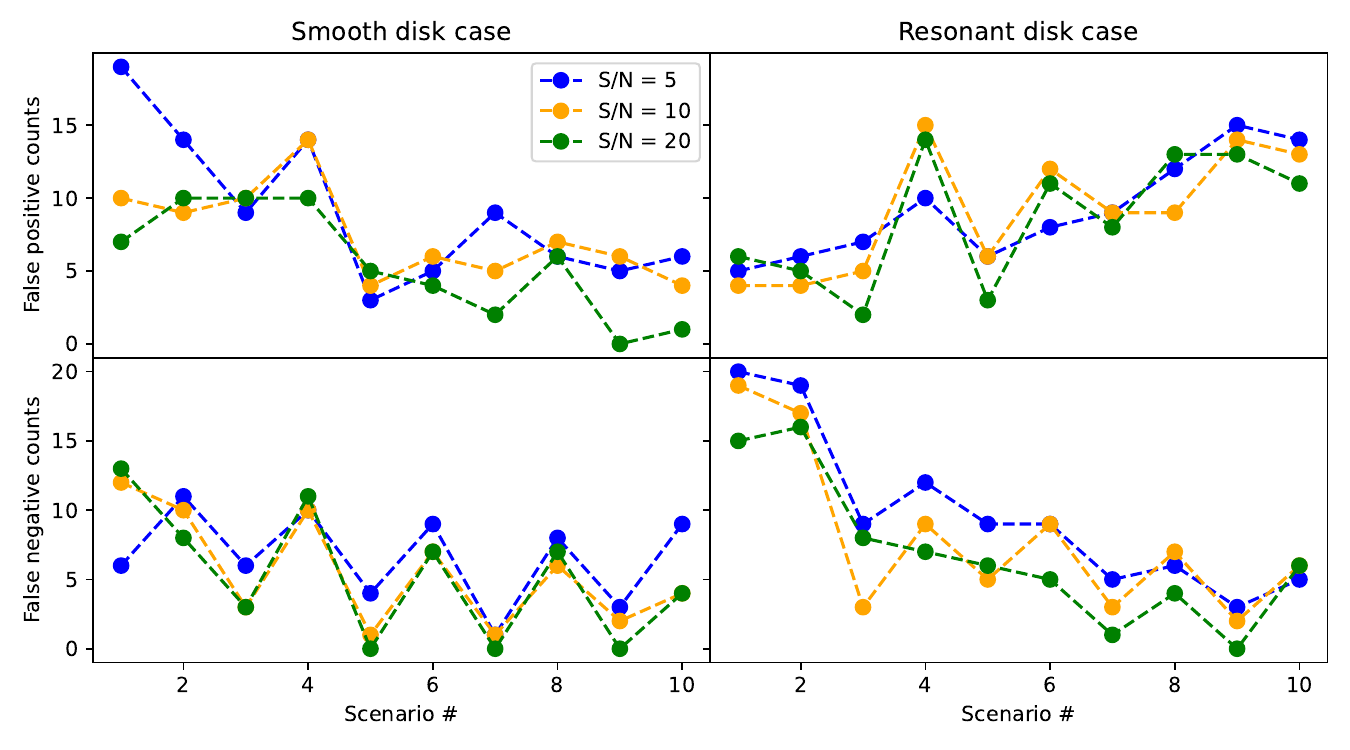}
		\caption{False positive and false negative counts for each astrophysical scenario considered for the smooth and resonant disk cases, reported by the Quartus Engineering team. \label{fig:false_positives_scenario}}
\end{figure}

It is also informative to tally the false positives and false negatives for each of the star-planet scenarios in Table~\ref{tab:scenarios}. As shown in Fig.~\ref{fig:false_positives_scenario}, the false positive counts decrease with larger Scenario \# in the smooth disk cases. This is probably related to the background/planet flux ratio assumed in the images (Table~\ref{tab:scenarios}). The planetary radii in scenarios 7 to 10 are larger than those in scenarios 5 and 6, which in turn are larger than those scenarios 3 and 4. Even though the assumed S/N for the inner planets is the same across these scenarios, for a fixed background, a larger planet implies a smaller background/planet flux ratio, resulting in less false positives. For the resonant disk cases, however, we observe a different trend in the false positive counts, where they increase for larger planets (Fig.~\ref{fig:false_positives_scenario}). This is likely because the dust clumps in the disk are created by the gravitational effects of the planets, and the larger the planet is, the larger the clumps will be \cite{stark2011transit}. The dust clumps are sometimes mistaken as planets and thus contribute to the false positive counts.

Meanwhile, the breakdown of the false negative counts by Scenario \# indicates the role of disk inclination (Fig.~\ref{fig:false_positives_scenario}). The false negative counts are higher for the high inclination cases (closer to ``edge-on'' in Scenarios 4, 6, 8, 10) and the false negative counts are quite low for the low inclination cases (closer to ``face-on'', scenarios 3, 5, 7, 9). This observation shows that, in an highly inclined planetary system, the planets may blend with and hide within the exozodiacal disk.
For Scenarios 1 and 2 ($\tau$~Ceti), which have three planets, we find that the planet 2 is generally not detected due to the high background/planet flux ratio (Table~\ref{tab:scenarios}), resulting in higher false negative counts. Similarly, Scenario 4 appears to score high counts of both false positives and false negatives. This scenario has small planets embedded in a highly inclined disk, making it a particularly challenging scenario for planet detection.

\subsection{Mississippi State University}

The more classical approach \cite{Lafreniere2007,Bowler2016} employed by the Mississippi State University (MSU) team allowed the analysis of a few images per scenarios. 
The first set of images analyzed by the MSU team do not have visible exozodi signals. 10 images from Scenario 1 and 2 ($\tau$~Ceti) have been identified in this category, and 14 planets have been reported with their positions and planet-star flux ratios. We confirm that all the planets reported are true detections, while there are additionally 18 planets in these images that are not reported. All inner planets in the 10 images are detected, and 4 of the outer planets are detected. We also find that the flux ratio reported is in average $\sim$30\% lower than the truth, suggesting that a background estimation is still necessary to correctly measure the planetary flux. On the same set of 10 images, the Quartus Engineering team reported 18 planets, 14 of which are true planets and 4 are false positive. The 14 true planets detected are identical between the two teams.

The second set presented by the MSU team uses multi-epoch subtraction. 12 images from 6 scenarios have been analyzed, and 10 planets have been reported. We also confirm that all the planets reported are true detections, while there are additionally 9 planets in these images that are not reported. The reported flux ratio is in average $\sim$25\% higher than the truth, again suggesting that the background estimation could be further optimized. Interestingly, for this second set of images, the Quartus Engineering team reported 25 planets of which only 5 are true planets. Because the Quartus Engineering team only uses single images in their analysis, this comparison suggests that combining multi-epochs images could better characterize the background for planet detection.

The last set presented by the MSU team uses the ZODIPIC software\cite{Kuchner2012} to model the disk for background subtraction. 8 images have been analyzed (4 from Scenario 1 and 4 from Scenario 2) and 10 planets have been reported. Of the planets reported, 9 are true detections and 1 is a false positive. There are 12 additional planets in these images. The reported flux ratio is in average $\sim$10\% larger than the truth, and as such, the method using a disk model appears to perform better than the previous two methods aforementioned. The inclination of the disk has been reported in this case: the inclination is reported to be 25$^{\circ}$ for the four images of Scenario 1 and 10$^{\circ}$ for the four images of Scenario 2, and the truth is 35$^{\circ}$. The reported values are compatible with the ``face on'' nature of the disk. For this third set of images, the Quartus Engineering team reported 20 planets, of which 12 are true detected and 8 are false positives.

While a meaningful comparison of multiple disk modeling software packages was beyond the scope of the modeling work, the MSU team carried out limited testing where they used alternative software packages, such as Diskmap \cite{Stolker2016}, to fit an astrophysical disk model to the images. One of the insights from these tests was that, when disk parameters are permitted to be unconstrained, the planet signals are often identified as point-like resonant disk structures. This highlights the challenges of determining an accurate computational model for disk subtraction, and indeed reflects the fundamental difficulty of distinguishing true planets from point-like disk structures.

Overall, by comparing the results from the two teams on the same images, the classical approach used by the MSU team tends to minimize the false positives, while the ICA-based approach used by the Quartus Engineering team reports more planets but also increases the number of false positives. This observation suggests that a combination of ICA-based background estimation and astrophysical disk modeling could maximize the planet detection rates while minimizing the false positives.

\section{Heuristic Model of Planet Detection}
\label{sec:heuristic}

\begin{figure}[!h]
\centering
	\includegraphics[width=\columnwidth]{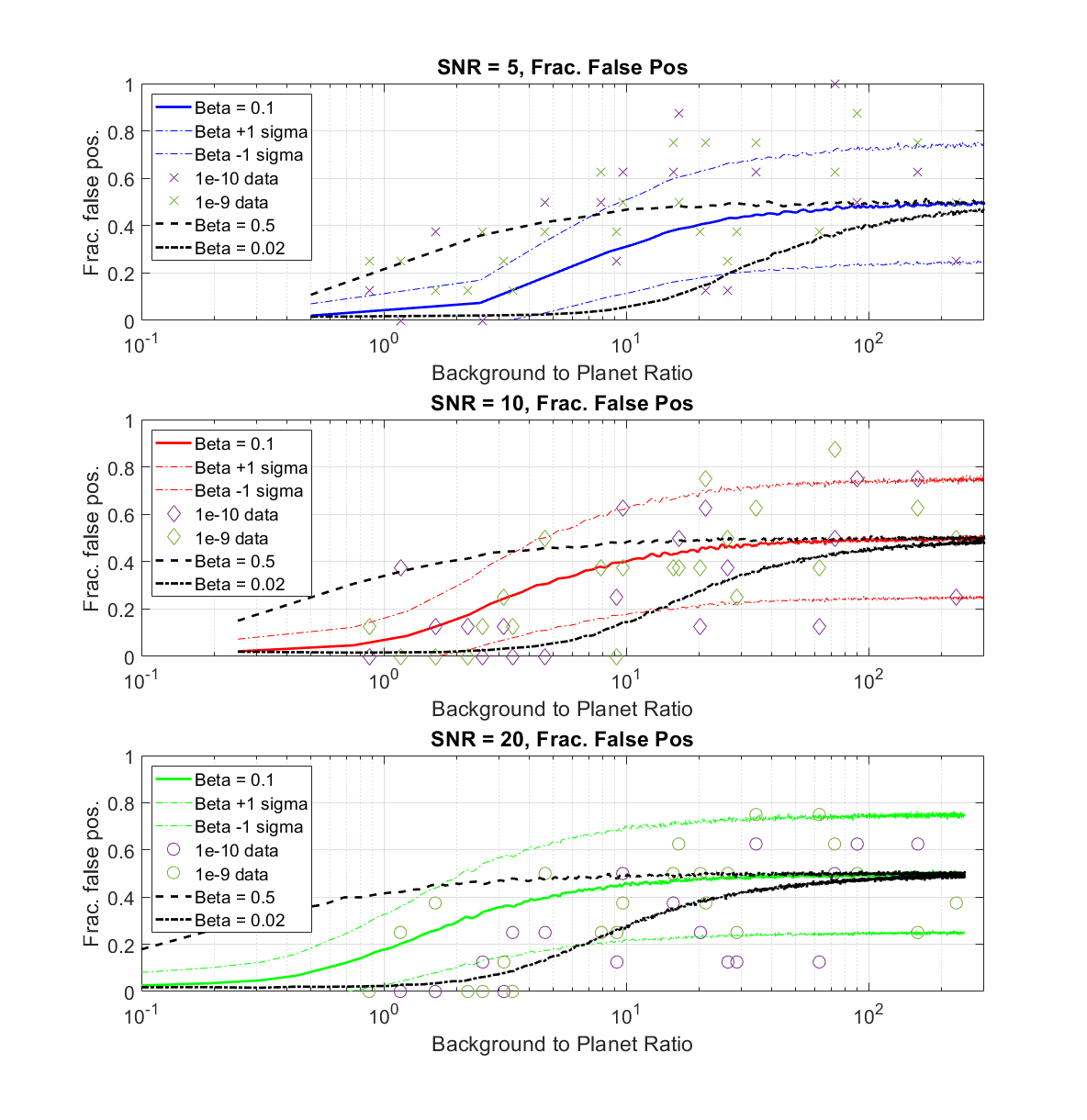}
  \caption{Quartus experimental results (symbols) and Monte-Carlo modeled fraction of false positives for idealized S/N = 5, 10, and 20, with background bias estimation standard deviation $\beta = 0.1$ (colored solid curves). Colored dashed curves show the $\pm 1\sigma$ error range assuming 8 trials at any background-to-planet ratio. The black curves show the predicted mean value for $\beta = 0.5$ and $\beta = 0.02$.  Curves are the result of 10,000 trials at each evaluated background-to-planet ratio.}
  \label{fig:frac_false_pos_stuart}
\end{figure}

\begin{figure}[!h]
\centering
	\includegraphics[width=\columnwidth]{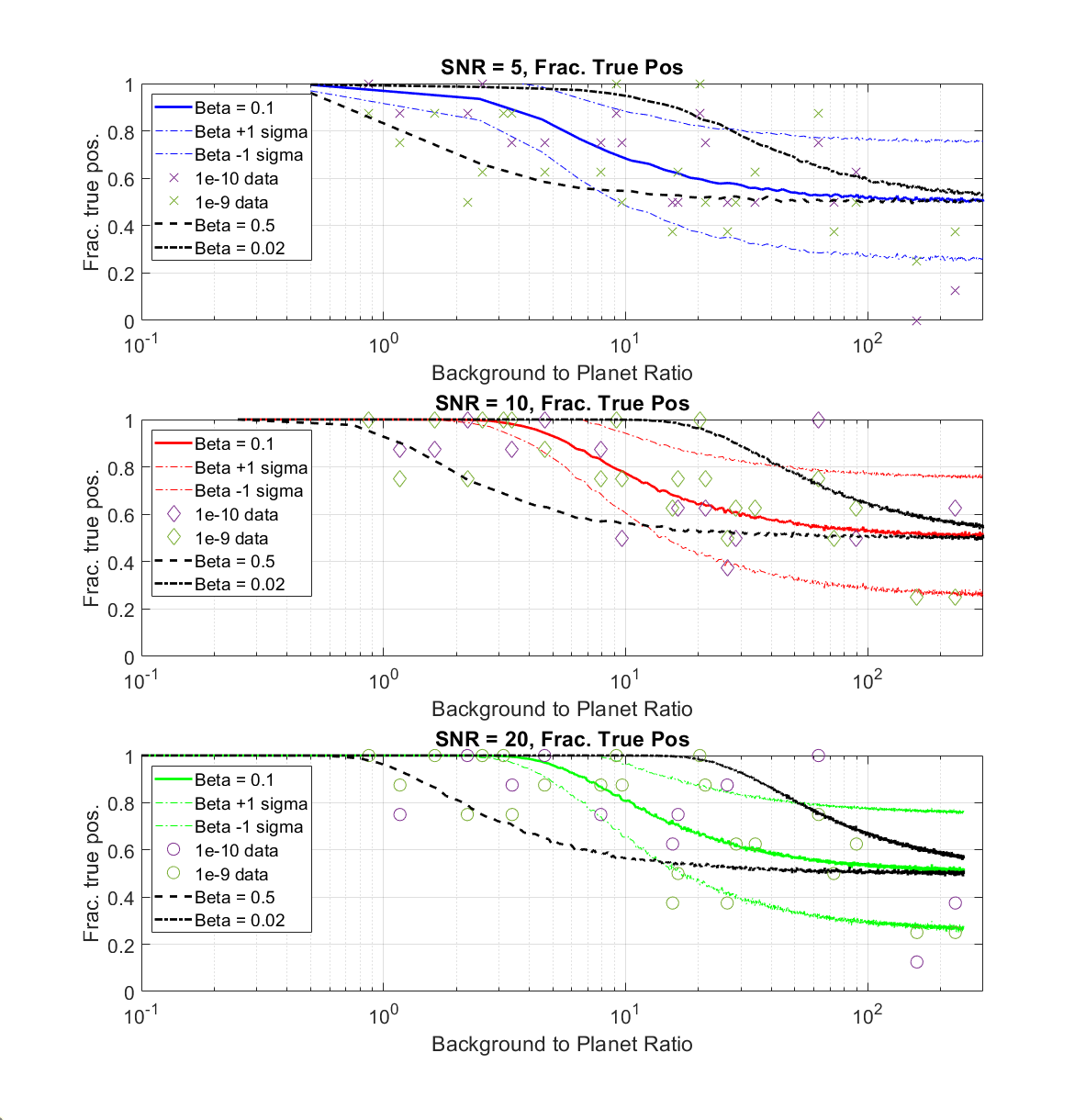}
  \caption{Quartus experimental results (symbols) and Monte-Carlo modeled fraction of true positives for idealized S/N = 5, 10, and 20, with background bias estimation standard deviation $\beta = 0.1$ (colored solid curves). Colored dashed curves show the $\pm 1\sigma$ error range assuming 8 trials at any background-to-planet ratio.The black curves show the predicted mean value for $\beta = 0.5$ and $\beta = 0.02$.  Curves are the result of 10,000 trials at each evaluated background-to-planet ratio.}
  \label{fig:frac_true_pos_stuart}
\end{figure}

All of the inner planets in the synthesized images in the data challenge have an idealized S/N $\ge5$. With an accurate estimate of the background, all of these planets should be detected. However, the Quartus Engineering pipeline applied to all the images resulted in a detection rate of $\sim70\%$ together with substantial number of false positives (Section~\ref{sec:tpfpfn}). Systematic background estimation errors result in the observed detection rates and false positives. Here we attempt to derive and constrain the underlying background estimation error, based on the planet detection statistics reported by the Quartus Engineering team. To do this, we devise a simple model to link the fractional background estimation error, $\beta$, to the planet detection (TP) and false positive (FP) rates. We focus on the inner planets because they formed the basis for the SNR = 5, 10, and 20 cases. The outer planet SNRs were not reported. Figs.~\ref{fig:frac_false_pos_stuart} and \ref{fig:frac_true_pos_stuart} use symbols to plot the reported inner planet FP and TP rates for the range of background-to-planet flux ratio in the simulated images. Here we break down the images into groups characterized by the background-to-planet ratio, and there are only 8 images in each group. The vertical spread of the data is a reflection of the small number of samples for each background-to-planet ratio.

The S/N of planet detection is expressed as\cite{Hu2021b}:
\begin{equation}\label{eq:noise_budget}
     S/N = \frac{N_P}{\sqrt{N_P + \alpha N_B + \beta^2 N_B^2}},
\end{equation}
where $N_P$ is the count from the planet, $N_B$ is the count from the background (which may include contributions of the leaked starlight, the solar glint, the exozodiacal dust, the local zodiacal dust, and the detector noises). The parameters $\alpha$ and $\beta$ result from the background estimation and subtraction (see Section~2.1 of Ref.~\citenum{Hu2021b}). 

We have set up a simple Monte-Carlo model to simulate the detection of exoplanets in the SEDC images. The model consists of 3 pixels: the first pixel (P1) has the flux from both the background and the planet. We use this pixel to estimate the TP rate, and the second and third pixels (P2 and P3) contain only the background. P2 is used to estimate the background, and P3 is used to estimate the FP rate.  

The planet is added to P1 with a count that is consistent with an idealized S/N of 5, 10, or 20 following the SEDC example. For each S/N, we adjust $N_P$ to generate a range of background-to-planet ratios by rearranging Eq.~\ref{eq:noise_budget} into Eq.~\ref{eq:Nback} with $\beta$=0, because we do not introduce a systematic bias in this pixel:
\begin{equation}
N_B = \frac{1}{\alpha}\bigg(\frac{N_P^2}{S/N^2}-N_P\bigg).
    \label{eq:Nback}
\end{equation}
P2 and P3 use the same mean count $N_B$. We explore the background-to-planet ratios that range between 0.25 and 250, covering both the planet-dominated and the background-dominated regimes.


With the mean counts established for each pixel, we generate 10,000 Poisson distributed noise realizations of each pixel and for each S/N and background-to-planet ratio. 
We account for systematic errors in the background estimation in the following way: in each realization $i$, the background estimate is equal to the counts in P2 multiplied by $1-\beta_i$, where $\beta_i$ is a normally distributed random number with a mean of zero and standard deviation of $\beta$. 
This background estimate is subtracted from P1 and P3 to form an estimate of the signal in these pixels. We then look at P1 and P3 to determine if a planet is present by comparing the residual counts to the standard deviation of the background.

The planet is detected when the background-subtracted signal is greater than a detection S/N threshold. We find that the final TP and FP rates are insensitive to the exact threshold value over a range $2 < {\rm S/N_{Thresh}} <4$. This is not surprising because all planets in the experiment are set to have an idealized S/N $>$ 5.

The results are, however, sensitive to $\beta$. Figs.~\ref{fig:frac_false_pos_stuart} and \ref{fig:frac_true_pos_stuart} use colored curves to show the M-C model predicted TP and FP rates for $\beta = 0.10$ and $\rm S/N_{Thresh}=3$ as a function of the background-to-planet ratio. The colored solid curves show the mean expected rates for each planet S/N case, while the colored dashed curves show the $\pm 1 \sigma$ rates assuming that eight experiments are conducted as was the case in SEDC. We also plot in black the mean model FP and TP rates for background estimation errors $\beta = 0.5$ and $\beta = 0.02$ (without accompanying $\pm 1 \sigma$ errors). Visually, the $\beta = 0.02$ curve underestimates FP and overestimates TP to at least background-to-planet ratios of 20, while the $\beta = 0.5$ curve overestimates FP and underestimates TP to background-to-planet ratio of $\sim 10$. The $\beta = 0.1$ curve and its $1\sigma$ error range capture most of the experimental results. Due to the small number of images at each background-to-planet ratio, the wide range of image conditions, e.g., symmetric versus non-symmetric exozodiacal disks, and the simplicity of our model compared to the Quartus Engineering analysis, we choose to not undertake a formal analysis to quantify the value of $\beta$ that best fits the data. But the simple heuristic model presented here indicates that significant background estimation errors exist in the range of $\sim10\%$.

\section{Discussion}
\label{sec:discussion}

The participating teams of SEDC have brought with them diverse expertise and approaches when analyzing the simulated high-contrast images obtained through a starshade. The Quartus Engineering team demonstrates that, with a fully non-parametrized and automatic algorithm, significant information on the planets and exozodiacal disks can be extracted from \textit{individual} images. The Mississippi State University team shows that, when necessary, classic astrophysical analysis (e.g., model-based disk characterization) could be applied to address false positives and false positives. The focus of the image post-processing approaches here is different from the approached developed and used by the high-contrast exoplanet imaging communities. This is because the current high-contrast imaging from the ground or using HST and JWST is limited by residual starlight through the coronagraph (i.e., the speckles), and the main objective of post-processing for planet detection is to remove this background through differential imaging or higher spectral resolution \cite{Lafreniere2007,soummer2012detection,Bowler2016,ruffio2024jwst}. With a starshade (or an advanced coronagraph being developed for the Habitable Worlds Observatory), the residual starlight no longer contributes significantly to the background noise; but for the purpose of finding Earth-sized planets, the planet signal is likely embedded in the background dominated by exozodiacal light. The results presented in this study are thus applicable broadly for anticipating the level of systematic errors and its impact on planet detection when the the background is dominated by exozodiacal light.

\subsection{Planet Detection}

This data challenge demonstrated the feasibility to detect planets and extract photometric measurements from starshade-assisted high-contrast observations. Using an independent component analysis to subtract the background, $\sim$70\% of the inner planets (close to the inner working angle, IWA) can be detected from individual images. Recall that we have applied an integration time that would result in idealized S/N of 5, 10, and 20 for the inner planets, and most of the non-detections come from the S/N = 5 images. This single-image planet recovery rate is significant as the regions of the image close to the IWA are often distorted by the edges of the starshade and contains the highest level of background from the exozodiacal dust disk. Therefore, our bona-fide tests have shown that it is possible to design effective algorithms to meaningfully subtract the background to expose the planetary signal. The results of the data challenge also suggest that it may be wise to design a planet search campaign that aims at an idealized (shot-noise-only) S/N of $\geq10$, as the planet recovery rate increases substantially from S/N = 5 to S/N = 10.

In addition, $\sim$40\% of the outer planets have been detected. While the regions are less impacted by instrumental noise, the light reflected by these planets is dimmer, and it is thus more challenging to extract them from the background. Therefore, a careful background subtraction is essential so that planets would not be removed by background subtraction (i.e., false negatives). 

The fact that we observe many false positives of the planet detection should be regarded as a cautionary tale. An inspection of the false positives image-by-image suggests that, the false positives found in the outer regions of the images are often associated with the clumpiness of the exozodiacal disk, while the false positives in the inner regions are more associated with the instrumental noise (leaked starlight, solar glint, and image distortion caused by the starshade) and incorrect background subtraction. The existence of false positives is not a surprise -- after all, the reported detections from the Quartus Engineering team are based on single images only. The ICA method for background estimation employed by the Quartus Engineering team provides a foundation for further improvement. Future studies should be carried out to determine the optimal ways to vet the false positives by astrophysics-based disk modeling, multi-wavelength image processing, and multi-epoch planet orbit reconstruction.

Finally, we want to emphasize the fundamental challenges to distinguish planets from point-like disk structures. As an example, the controversy regarding Fomalhaut~b, which resides in one of the most easy-to-view and heavily studied debris disks, highlights the potential challenges to distinguish true planets from disk structures, including collision-induced structures. Even after more than a decade of multi-wavelength investigations, disk modelling, and orbital monitoring, it is still unclear whether the object is a planet or dust clumps that form due to the collision of planetary bodies \cite{Wyatt2004,Kalas2008}. The SEDC did not include the effects of planetary collisions, but the substantial rates of false positives associated with the resonant disk structures suggest that distinguishing point-like disk structures should be a focus of future studies of post processing and planet detection in the context of spaceborne high-contrast imaging.

\subsection{Systematic Errors in Background Estimation}

The SEDC gives us the possibility to derive an empirical estimate of the achievable quality of background subtraction, based on the reported detections of planets, false positives and negatives, and the reported photometric measurements. We emphasize that the participating teams had no knowledge of the truths, and so the SEDC provides a rare ``blind'' test of the background estimation and subtraction. In Section~\ref{sec:heuristic}, we show that the outcomes of the Quartus Engineering analysis imply a mean background estimation error of $\sim 10\%$. This background estimation error explains most of the false positives and false negatives found.
The background estimation error also results in an irreducible noise floor: $S/N\rightarrow N_P/\beta N_B$ when the integration time goes to infinity, according to Eq. (\ref{eq:noise_budget}). This study recommends $\beta=0.1$ to be included in a more realistic noise budget when anticipating the science returns of future direct imaging instruments working on the exozodi-limited regime. The concept of the background estimation error has also been recognized in the context of exoplanet direct imaging using a coronagraph \cite{nemati2020method,mennesson2024current}, and in addition to the exozodiacal light, the stability of the residual starlight through a coronagraph would also contribute to the imperfect background calibration.

The finding of significant background estimation error here appears to be inconsistent with the recent suggestions that the exozodiacal disk and other background signals in high-contrast images can be subtracted to close to the photon-noise limit (i.e., $\beta\sim0$ in our notation) over a wide range of conditions \cite{kammerer2022simulating,currie2023mitigating}. Interestingly, both Refs.~\citenum{kammerer2022simulating,currie2023mitigating} and SEDC find that the background estimation for the purpose of planet detection becomes more difficult when the exozodi level and system inclination increase. The new revelation here is that even with a relatively low background-to-planet ratio of 10, and a high idealized S/N of 10, the background estimation error can still result in substantial false positives and negatives, as well as planet flux estimation errors. It is important to interpret our finding in the context of the method used for background estimation, which was a non-parameteric algorithm without modeling the underlying astrophysical phenomena. It is entirely possible that the performance of the algorithm can be improved in the future, for example by folding in the exozodiacal disk models \cite{stark2011transit}, and the implied $\beta$ value could be smaller than what we currently find. For example, using convotional neural networks on the same set of synthesized images could result in substantial improvement in the planet recovery rate, while generalization remains challenging \cite{ahmed2023exoplanet}. As all the synthesized images and the true values are available from the SEDC webpage, we encourage the community to leverage this unique dataset and continue exploring image processing techniques for background estimation and planet detection in the exozodi-limited regime.

\section{Conclusion and prospects}
\label{sec:conclusion}

In this paper, we present the outcome of the Starshade Exoplanet Data Challenge, for which the design and rationale have been described in Ref.~\citenum{Hu2021}. We describe the methods taken by the two participating teams to analyze the synthetic images. One method focuses on the automation of the program to analyze large quantity of images and employs a non-parametric modeling and subtraction of the background, while the other method uses a classic approach with an attempt to derive a physics-based description of the exozodiacal disk in the image. 

From the large ensemble study of the synthetic images and their analysis, we demonstrate the feasibility of detecting small exoplanets embedded in the exozodiacal disk from individual images. We show that 70\% of the inner planets can be detected, and using an integration time that yields an idealized S/N of $\geq10$ would further improve the detection rate. We also show that about half of the outer planets can be detected. Using a non-parametric background model, however, would also yield false positives, and we find that many of the false positives are associated with the assumed dust clumps of exozodiacal disks in orbital resonance with the planets. Further understanding of the structure of the exozodiacal disks, or a combination of non-parameteric models and physics-based models, could help reduce the number of false positives. In general, we find that planet detection is more challenging in systems that have a dense or highly inclined exozodiacal disk.

By comparing a heuristic model to the reported false positives/negatives and planetary fluxes in this large ensemble study, we derive an empirical estimate of the residual background, and find the mean background estimation error to be $10\%$ in the ICA-based, non-parameteric algorithm. Note that the algorithm was not trained in any way as the truths were not disclosed to the participating teams before their submission of the results. While further improvement of background estimation algorithms holds the promise to reduce the background estimation error, it is important to note that this systematic error term is in addition to the photon noise in the noise budget (Eq.~\ref{eq:noise_budget}), leads to an irreducible noise floor, and should be considered when estimating the science capability of future missions.

High-contrast imaging will be the main method to unveil small planets in the habitable zones of Sun-like stars. The insight into the noise budget and the image post-processing gained in SEDC will be useful to design the technology and system architecture needed to detect and characterize these planets. As the dominant noise source in the synthetic images in this study is exozodiacal light and not residual starlight (or any other starshade-specific noise), the insight learned here is also applicable to exoplanet direct imaging using a coronagraph that exquisitely suppresses the starlight and enables exozodi-limited imaging. The images synthesized by the SEDC will be a good test ground for new methods of image processing; for example, with the advance of computer vision, one may expect the application of methods based on neural networks to reveal the planets in the images. The community data challenge is, again, proven to be a very effective way to advance our understanding of the science capability of space systems.

\section*{Code and Data Availability}

The synthesized images, the true values, and the final reports of the participating teams are available at NASA Exoplanet Exploration Program's website (https://exoplanets.nasa.gov/exep/technology/starshade-data-challenge). The image analysis code of Quartus Engineering is available at Github (https://github.com/bdunne6/Exoplanet-Detection-Dev).

\section*{Acknowledgments}

The research was carried out at the Jet Propulsion Laboratory, California Institute of Technology, under a contract with the National Aeronautics and Space Administration (80NM0018D0004). The work performed by A.N. and J.C. was also supported by the South Carolina Space Grant Consortium.

\end{spacing}
\end{document}